\input{psfig}

\documentstyle[prl,aps,twocolumn]{revtex}
\begin{document}

\title{\vspace*{-0.5cm}
\hspace*{\fill}{\normalsize LA-UR-97-324} \\[1.5ex]
The Linear Correlation Coefficient vs. the
Cross Term in Bose-Einstein Correlations}
\author{
B.R. Schlei${}^1$\thanks{E. Mail: schlei@t2.LANL.gov}{\ },
D. Strottman${}^1$\thanks{E. Mail: dds@LANL.gov}{\ }, and
N. Xu${}^2$\thanks{Present address: MS 50A, 1 Cyclotron Rd., LBNL.}{\ }
\\[1.5ex]
{\it ${}^1$Theoretical Division, Los Alamos National Laboratory,
Los Alamos, NM 87545, USA}\\
{\it ${}^2$P-25, Los Alamos National Laboratory,
Los Alamos, NM 87545, USA}
}
\date{\today}
\maketitle

\begin{abstract}
We investigate the nature of the new cross term for Gaussian
parameterizations of Bose-Einstein correlations of identical particles
emitted from purely chaotic hadron sources formed by relativistic
heavy ion collisions. We find that this additional parameter in the
so-called Bertsch parameterization can be expressed in terms of a
linear ``out-longitudinal'' correlation coefficient for emission of
bosons and two already known ``radius'' parameters, $R_l$ and $R_o$. The
linear correlation coefficient is of kinematical nature and can be used 
to determine the widths of longitudinal momentum distributions.\\
\end{abstract}

\hbadness=10000

In the ongoing search for the quark-gluon plasma (QGP) one is especially
interested in the volumes and the lifetimes of the hot and dense zones of
nuclear matter, {\it i.e.}, the fireballs that are generated in relativistic
heavy-ion collisions.  A tool to measure space-time extensions
\cite{bernd11} of the fireball is given through the Hanbury-Brown/Twiss
(HBT) interferometry \cite{GGLP}, which provides the measurement of
Bose-Einstein correlations (BEC) \cite{boal}.\\ 
BEC functions of identical bosons are quantum-statistical observables which 
contain information about ten quantities \cite{michael} which characterize 
a hadron source: lifetimes, longitudinal and transverse extensions of
the chaotic and the coherent source, temporal and spatial
(longitudinal and transverse) coherence lengths and the chaoticity. In
the case of pion interferometry it has been argued that a possible
coherent subcomponent of the hadron source is almost unobservable
\cite{bernd3}, because resonance decays contribute strongly to
BEC. Many parameterizations of experimental BEC functions are based on
a neglect of a coherent component thereby resulting in the description
of the hadron-emitting source with only three quantities, ({\it e.g.},
lifetime, longitudinal and transverse radii of the purely chaotic
hadron source).\\ 
For heavy-ion collisions a very popular parameterization of purely chaotic 
BEC functions is provided through the so-called Bertsch parameterization
\cite{bertsch} wherein a two-particle BEC function is parameterized in terms
of the three momentum-dependent ``radius'' parameters $R_l(\vec{K})$,
$R_s(\vec{K})$, and $R_o(\vec{K})$. Here, $\vec{K} \equiv 
\textstyle{\frac{1}{2}}(\vec{k}_1 + \vec{k}_2)$ is the average momentum of
two identical bosons which are emitted with their individual momenta,
$\vec{k}_1$ and $\vec{k}_2$, respectively. In the HBT Cartesian coordinate system
the ``longitudinal'' or $z$ direction (subscript $l$) is parallel to the beam, 
the ``out'' or $x$ direction (subscript $o$) is parallel to the component of 
$\vec{K}$ which is perpendicular to the beam, and the ``side'' or $y$ direction 
(subscript $s$) is the remaining  direction. In ref. \cite{chapman} it was shown 
that even in the case of a purely chaotic hadron source, three quantities are 
not adequate to fully parameterize the system under consideration. {\it E.g.}, a 
cylindrically symmetric expanding fireball has in general to be described with 
an additional ``cross-term'' radius\footnote{Experimentally, a ``cross-term'' 
has been observed, {\it e.g.}, by the NA35/NA49 Collaborations \cite{qm95na49}. 
}, $R_{ol}(\vec{K})$.\\
The generalized (Gaussian) Bertsch parameterization of a purely chaotic
two-particle Bose-Einstein correlation function is given through \cite{chapman}
\begin{eqnarray}
C_2(\vec{k}_1,\vec{k}_2)\:=\: 1&+&\lambda(\vec{K})\cdot\exp
[-\:q_l^2 R_l^2(\vec{K})\:-\:q_o^2 R_o^2(\vec{K})\:
\nonumber\\
&&-\:q_s^2 R_s^2(\vec{K})\:
-2\:q_o q_l R_{ol}^2(\vec{K})] \:.
\nonumber\\
\label{eq:chapman}
\end{eqnarray}

In eq. (\ref{eq:chapman}) the $q_i$ ($i = l, s, o$) refer to the
components of the momentum difference $\vec{q} = \vec{k}_1 - \vec{k}_2$, and
$\lambda(\vec{K})$ is the momentum dependent incoherence factor which
accounts for reductions of the BEC due to long-lived resonances \cite{bernd3}
and averaging due to phase-space, respectively.  

In a physical interpretation one is troubled by the fact that one and the
same two-particle BEC function is described in case (a) of the generalized
Bertsch parameterization through four ``radius'' parameters ($R_l$,
$R_s$, $R_o$, $R_{ol}$), and (b) in case of the Yano-Koonin-Podgoretski\u{\i} 
(YKP) parameterization \cite{heinzYKP} through three ``radius''  
parameters ($R_0$,
$R_l$, $R_\perp$) and one kinematic quantity, the Yano-Koonin
velocity ($v$). It is the purpose of this paper to take a closer look at the
recently reported new cross term, $R_{ol}$, and to investigate its
nature.

Before we begin our analysis, we recall that the four ``radius''
parameters of the generalized Bertsch parameterization can be expressed in 
terms of variances of the (boosted) components of the space-time points 
$x^\nu = (t, x, y, z) \in \Sigma$ ($\Sigma$ is the freeze-out hypersurface) 
and covariances between the components of $x^\nu$ ({\it cf.} ref. 
\cite{bernd11})

\parbox{3.0cm}{
\begin{eqnarray}
&&R_s^2(\vec{K})\: \approx \: \sigma^2_y \:,
\nonumber\\
&&R_l^2(\vec{K})\: \approx \: \sigma^2_{z\:-\:\beta_\parallel t} \:,
\nonumber
\end{eqnarray}}
\hfill\parbox{5.0cm}{
\begin{eqnarray}
&&R_o^2(\vec{K})\: \approx \: \sigma^2_{x\:-\:\beta_\perp t}\:,
\label{eq:radii}
\\
&&R_{ol}^2(\vec{K})\: \approx \: \sigma_{\:x\:-\:\beta_\perp 
t,\:z\:-\:\beta_\parallel t}\:.
\nonumber
\end{eqnarray}}

In eq. (\ref{eq:radii}) we use $\beta_i=K_i/E_K$ ($i = \parallel, \perp$), where
$E_K=\sqrt{m^2+\vec{K}^2}$.

Let us now consider a quite general mathematical representation of a
two-dimensional Gaussian, $f(q_l,q_o)$, in the two variables $q_l$ and $q_o$
with fixed parameters $R_l$, $R_o$, and $\lambda$
\begin{eqnarray}
f(q_l,q_o)\:=\:1&+&\lambda\cdot\exp[-\:q_l^2 R_l^2
\:-\:q_o^2 R_o^2\:
\nonumber\\
&&+\:2\:q_o q_l R_o R_l\:\rho_{ol}]\:,
\label{eq:func2}
\end{eqnarray}

where $\rho_{ol}$ is the ``linear correlation  coefficient'' which
gives the strength of linear correlation between the two quantities 
$q_l R_l$ and $q_o R_o$. A comparison between eq. (\ref{eq:chapman}) 
 and eq. (\ref{eq:func2}) yields for the ``cross-term'' (while ignoring the 
``side-term'' without loss of generality)
\begin{equation}
R_{ol}^2(\vec{K})\:\equiv\: -\:\rho_{ol}(\vec{K})\:\cdot\:
R_o(\vec{K})\:R_l(\vec{K})\:,
\label{eq:rcross}
\end{equation}

or in terms of the variances and covariances for the (boosted) ``longitudinal'' 
and ``out'' components of $x^\nu$
\begin{equation}
\rho_{ol}(\vec{K})\:\equiv\:-\:\frac{R_{ol}^2(\vec{K})}
{R_o(\vec{K})\:R_l(\vec{K})}
\:\approx\:-\:
\frac{\sigma_{x-\beta_\perp t,\:z-\beta_\parallel t}}{
\sqrt{\sigma_{x-\beta_\perp t}^2\:\sigma_{z-\beta_\parallel t}^2}}\:,
\label{eq:rho}
\end{equation}

with $|\rho_{ol}(\vec{K})|\:\leq\:1\quad\forall\:\vec{K}$, representing the 
Cauchy-Schwarz inequality relation for $\rho_{ol}$.

Thus the ``cross-term'' radius parameter, $R_{ol}$ in the Bertsch
parameterization can be expressed in terms of a linear ``out-longitudinal''
correlation coefficient, $\rho_{ol}$, for emission of bosons and the
two already known ``radius'' parameters, $R_l$ and $R_o$.

In order to illustrate the nature of the cross-term, in Fig.1  we plot the
pion rapidity distribution (top) and the scatter plot
$q_l$ vs. $q_o$ (bottom). These plots are the results of the
microscopic transport model RQMD \cite{rqmd21} calculations for 200 AGeV
S+S central collisions. All plots are generated in the nucleon-nucleon
center of mass system; the rapidity windows are labeled in the figure.

In the mid-rapidity window no correlation between $q_l$ and $q_o$
is seen while sizable correlations are found in other rapidity
windows. In addition the strength of the correlation depends on the
rapidity for which one constructs the scatter plot\footnote{ Note
that in case of a purely Gaussian distribution a non-vanishing
$\rho_{ol}$ is represented by a tilted ellipsoid. However, inspecting
Fig. 1 we found a non-ellipsoid shape in the scatter plots. This
feature indicates non-Gaussian distributions in real collisions.
This will not alter our main conclusion in the text.}. Qualitatively,
the strength of the correlation is consistent with the NA35/NA49
result
\cite{qm95na49}. No correlation is seen between $q_l$ and
$q_s$ or $q_s$ and $q_o$ in this model calculation. As discussed
later in more detail, this longitudinal momentum dependence implies
that the cross-term is kinematic in nature. However, as can be seen
in the figure this dependence could be used to determine the rapidity
width of the particles under study.

For the sake of further illustration we consider results of a hydrodynamical
analysis \cite{bernd7,bernd8} of 160 AGeV $Pb+Pb$ (central) collisions using
the HYLANDER code\cite{udo}. In the following we discuss the fireball
emission of (identical) $K^-K^-$ pairs because in ref. \cite{bernd11} it has
been argued that for $K^-K^-$ BEC functions their numerical calculation and
subsequent fit can be replaced by the direct evaluation of eqs.
(\ref{eq:radii}) in order to determine the four ``radius'' parameters of the
generalized Bertsch parameterization. We perform our analysis for average, 
transverse kaon pair momenta $K_\perp = 1.5$ GeV/c (an analysis for different 
average transverse kaon pair momenta, $K_\perp \neq 0$, leads qualitatively to 
the same results).

We now introduce the strengths of linear correlation among $t$, $x$, and $z$
\begin{equation}
\rho_{t,x}\:\equiv\:\frac{\sigma_{t,x}}{\sqrt{\sigma_t^2\:\sigma_x^2}}\:,
\:\:\rho_{t,z}\:\equiv\:\frac{\sigma_{t,z}}{\sqrt{\sigma_t^2\:\sigma_z^2}}\:,
\:\:\rho_{x,z}\:\equiv\:\frac{\sigma_{x,z}}{\sqrt{\sigma_x^2\:\sigma_z^2}}\:.
\label{eq:rhoij}
\end{equation}

From eqs. (\ref{eq:radii}) and (\ref{eq:rhoij}) we are now able to rewrite 
$R_l$, $R_o$, and $R_{ol}$
\begin{eqnarray}
R_o^2(\vec{K}) &\approx&
\sigma^2_x\:+\:\beta_\perp^2 \sigma_t^2\:-\:2\:\beta_\perp\sigma_t\sigma_x
\:\rho_{t,x}\:,
\label{eq:radii2}
\\
R_l^2(\vec{K}) &\approx&
\sigma^2_z\:+\:\beta_\parallel^2 \sigma_t^2\:-\:2\:\beta_\parallel
\sigma_t\sigma_z\:\rho_{t,z}\:,
\nonumber\\
R_{ol}^2(\vec{K}) &\approx& \sigma_x\sigma_z\:\rho_{x,z}\:
+\:\beta_\parallel\beta_\perp\sigma_t^2\:
\nonumber\\
&&-\:\beta_\parallel\sigma_t\sigma_x\:\rho_{t,x}
\:-\:\beta_\perp\sigma_t\sigma_z\:\rho_{t,z}\:.
\nonumber
\end{eqnarray}

In Fig. 2 we show the linear correlation coefficients due to eqs.
(\ref{eq:rhoij}) and (\ref{eq:rho}) for emission of negative kaons as a
function of the kaon-pair rapidity, $y_K$, at $K_\perp = 1.5$ GeV/c.  The
results have been obtained for $Pb+Pb$ at 160 AGeV with  a rapidity,
$y_\Delta = 0.90$, at the edge of the initial central fireball  ({\it cf.}
ref. \cite{bernd8} for more details). Fig. 2 is divided into three ranges of
the kaon-pair rapidity, $y_K \leq -1.4$, $|y_K| < 1.4$ ,and $1.4 \leq y_K$, 
which correspond to the freeze-out hypersurface emission zones SWR, NTR, SWR, 
respectively. SWR denotes the ``simple wave region'' which is given by the 
original Riemann rarefaction waves, and NTR denotes the ``non-trivial region'',
where two rarefaction waves overlap ({\it cf.} \cite{hydro} and also Fig. 3). 
I.e., kaon pairs with $|y_K| < 1.4$ are mostly emitted from the NTR, 
while kaon pairs with $1.4 < |y_K|$ are mostly emitted from the SWR.

Fig. 2(a) shows that there is an almost perfect linear correlation between
the emission times $t$ and the transverse emission coordinates $x$ (the
longitudinal emission coordinates $z$) in the NTR (SWR) of the freeze-out
hypersurface. We have $|\rho_{t, x}| \approx 0.85$ (dotted line) in the NTR, and
$|\rho_{t, z}| \approx 1.0$ (dashed line) in the SWR.  For $|y_K|
\longrightarrow \infty$ we get $|\rho_{t, x}| \longrightarrow 0$,  and
because of symmetry reasons we have $\rho_{t, z}(y_K=0) = 0$. The linear
correlation coefficient $\rho_{x, z}$ (solid line) reaches its extrema at the
transition zones of NTR and SWR.

Fig. 2(b) shows that by coincidence $\rho_{x,z}\approx\rho_{t,x}\cdot\rho_{t,z}$, 
but this relation is in general not true. Fig. 2(c) shows that there is a direct 
proportionality between the (internal) linear correlation coefficient, 
$\rho_{x, z}$ (solid line), and the measurable $\rho_{ol}$ (dashed line):
\begin{equation}
\rho_{ol}(\vec{K})\:\propto\:\rho_{x,z}\:.
\label{eq:radii4}
\end{equation}
Because of this proportionality $\rho_{ol}$ reflects the strength of
linear correlations between transverse and longitudinal emission
coordinates of the freeze-out hypersurface. A negative value for
$\rho_{ol}$ at positive average kaon-pair rapidities indicates that
for larger longitudinal emission positions, the transverse emission
positions become smaller. Conversely, a positive value for
$\rho_{ol}$ at negative average kaon-pair rapidities indicates that
for larger longitudinal emission positions (in the negative direction)
the transverse emission positions also become smaller. Hence,
$\rho_{ol}$ enables us to measure finite-size effects of fireballs in
rapidity space.

In Fig. 3 we show (a) the rapidity spectra, $dN/dy$, (b) linear correlation
coefficients, $\rho_{ol}$, and (c) freeze-out hypersurfaces
$\Sigma(r=0)$ for emission of negative kaons for $Pb+Pb$ at 160 AGeV.  The
results have been obtained with different rapidities, $y_\Delta$, at the
edge of the initial central fireballs. The parameter $y_\Delta$ controls 
the strength of the initial rapidity field of the initial fireball in the 
hydrodynamical model. A larger value for $y_\Delta$ results in a larger initial 
longitudinal rapidity field of the
fluid, and subsequently at freeze-out it leads to rapidity spectra of larger 
widths ({\it cf.} Fig. 3(a)).  Fig. 3(c) shows the corresponding freeze-out 
hypersurfaces $\Sigma(r=0)$. Consistent with expectation, a larger value for 
$y_\Delta$ results in more longitudinally extended freeze-out hypersurfaces. 
Furthermore, we observe that with increasing $y_\Delta$, the NTR of the 
freeze-out hypersurfaces becomes more and more hyperbola shaped (like the 
scaling solution for Bjorken initial conditions \cite{bjorken}).

We stress, as shown with Fig. 3(b), that the Full Widths at Half Maximum of
the rapidity spectra, $dN/dy$, of negative kaons nearly coincide with the
positions of extrema of the corresponding linear  ``out-longitudinal''
correlation coefficients, $\rho_{ol}(\vec{K})$. We also observe that a
more hyperbola shaped freeze-out hypersurface results in smaller absolute
values for $\rho_{ol}$.

After introducing the linear correlation coefficient, $\rho_{ol}$, the
difficulties of interpretation encountered above no longer persist, if one
uses in case of the generalized Bertsch parameterization the three
``radius'' parameters, $R_l$, $R_s$, and $R_o$, and the
kinematical quantity $\rho_{ol}$. In general, if one analyzes a heavy
ion collision which has no azimuthal symmetry, further (measurable)
linear correlation coefficients come into existence. In case of the
generalized (Gaussian) Bertsch parameterization, one  would use for a fit of
a two-particle BEC function ($i,j = l, s, o$)
\begin{eqnarray}
C(\vec{K},\vec{q}\:) &=& 1\:+\:\lambda(\vec{K})
\label{eq:general}
\\
&&\times\:\exp\left[-\: \sum_{ij} \: q_i R_i(\vec{K}) 
\:q_j R_j(\vec{K})\:\tilde{\rho}_{ij}(\vec{K})\right]\:,
\nonumber
\end{eqnarray}
with
\begin{eqnarray}
\tilde{\rho}_{ij}(\vec{K})&\equiv&
\left(
\begin{array}{ccc}
1 & -\:\rho_{os} & -\:\rho_{ol} \\
-\:\rho_{os} & 1 & -\:\rho_{sl} \\
-\:\rho_{ol} & -\:\rho_{sl} & 1
\end{array}
\right)
\label{eq:matrix}\\
&&\nonumber\\
&=&-\:\frac{1}{2\:R_i(\vec{K})\:R_j(\vec{K})}\:
\frac{\partial^2\:C(\vec{K},\vec{q}\:)}{\partial q_i\:\partial q_j}
\bigg|_{\vec{q}\:=\:0}
\label{eq:deriv}\\
&&\nonumber\\
&\approx&
\frac{\sigma_{x_i-\beta_i t,\:x_j-\beta_j t}}{
\sqrt{\sigma_{x_i-\beta_i t}^2\:\sigma_{x_j-\beta_j t}^2}}\quad,
\label{eq:varco}
\end{eqnarray}
and $|\rho_{ij}(\vec{K})| \leq 1 \quad\forall \vec{K}$. In eq.
(\ref{eq:varco}), we have used the notation $x_l\equiv z$,
$x_o\equiv x$, $x_s\equiv y$, $\beta_l\equiv\beta_\parallel$,
$\beta_o\equiv\beta_\perp$, and $\beta_s\equiv 0$.

To summarize, we have investigated the cross term for Gaussian
parameterizations of Bose-Einstein correlation functions of purely
chaotic hadron sources. We find that at relativistic energies this
additional parameter in the so-called Bertsch parameterization can be
expressed in terms of a linear ``out-longitudinal'' correlation
coefficient, $\rho_{ol}(\vec{K})$, for emission of bosons and the two
already known ``radius'' parameters, $R_l$ and $R_o$. The
correlation coefficient provides an additional tool for determining
the particle rapidity distributions.\\

We thank Dr. H. Sorge for providing us the RQMD code. We are grateful
for many instructive discussions with Drs. U. Heinz, E.V. Shuryak,
H. Sorge, and Yu.M. Sinyukov. This work has been supported by the
U.S. Department of Energy.

\vspace*{6.0cm}
\begin{figure}
\caption{Rapidity spectrum of negative pions in arbitrary units
for $S+S$ at 200 AGeV from RQMD (version 2.1). Pion pairs have been
formed in different rapidity bins and their corresponding correlations
between $q_o$ and $q_l$ are shown as scatter plots. The different
colors in these plots indicate different relative strengths of
the correlation functions (seven equidistant regions). The correlation
functions take their maximum values (near to 2) in the center of each
scatter plot. The large red areas indicate a minimum correlation
strength near to 1.}
\label{fg:fig1}
 \end{figure}

\newpage
\vspace*{6.5cm}
\begin{figure}
\caption{Linear correlation coefficients calculated from
HYLANDER, for emission of negative kaons for $Pb+Pb$ at 160 AGeV as a
function of the kaon-pair rapidity, $y_K$, at $K_\perp$ = 1.5 GeV/c.
(a) $\rho_{t, x}$ (dotted line), $\rho_{t, z}$ (dashed line),
$\rho_{x, z}$ (solid line); (b) the product $\rho_{t, x}\cdot\rho_{t,
z}$ (dashed-dotted line), $\rho_{x, z}$ (solid line); (c) the
observable $\rho_{ol}$ (dashed line), and $\rho_{x, z}$ (solid line).}
\label{fg:fig2}
 \end{figure}

\vspace*{10.0cm}
\begin{figure}
\caption{(a) Rapidity spectra, $dN/dy$, (b) linear correlation
coefficients, $\rho_{ol}$, at $K_\perp$ = 1.5 GeV/c and (c)
freeze-out hypersurfaces $\Sigma(r=0)$ for emission of negative kaons
for $Pb+Pb$ at 160 AGeV. The four line styles correspond to four different 
rapidities, $y_\Delta$, at the edge of the initial central fireballs.}
\label{fg:fig3}
 \end{figure}


\end{document}